\documentclass[11pt, a4paper]{article}
\usepackage{float}
\usepackage{natbib}
\usepackage[utf8]{inputenc}
\usepackage[margin=1in]{geometry}
\usepackage{amsmath, amssymb, amsfonts}
\usepackage{bm}
\usepackage{siunitx}
\usepackage{booktabs}
\usepackage{graphicx}

\usepackage{authblk}

\usepackage{hyperref}
\hypersetup{colorlinks=true, linkcolor=red, citecolor=blue, urlcolor=black, breaklinks=true}
\usepackage[capitalize,nameinlink]{cleveref}

\title{\textbf{ForestIR: Physics-Informed Forest Sound Simulation for Array-Based Bioacoustic Remote Sensing}\vspace{1.5em}}

\author[1]{Xin Shen}
\author[1,2]{Jennifer N. Kampe}
\author[1]{Changwoo J. Lee}
\author[1]{Braden Scherting}
\author[3]{Panu Somervuo}
\author[4]{Ari Lehtiö}
\author[4]{Sandro von Brandenburg}
\author[2,5]{Ossi Nokelainen}
\author[2]{Otso Ovaskainen}
\author[1]{David B. Dunson}

\affil[1]{Department of Statistical Science, Duke University, Durham, NC, USA}
\affil[2]{Department of Biological and Environmental Science, University of Jyväskylä, Jyväskylä, Finland}
\affil[3]{Organismal and Evolutionary Biology Research Programme, Faculty of Biological and Environmental Sciences, University of Helsinki, Helsinki, Finland}
\affil[4]{Digital Services, University of Jyväskylä, Jyväskylä, Finland}
\affil[5]{Open Science Centre, University of Jyväskylä, Jyväskylä, Finland}
\date{}

\begin{document}

\maketitle

\newcommand\blfootnote[1]{%
  \begingroup
  \renewcommand\thefootnote{}\footnote{#1}%
  \addtocounter{footnote}{-1}%
  \endgroup
}
\blfootnote{Correspondence: xin.shen@duke.edu (Xin Shen). Senior authors: otso.t.ovaskainen@jyu.fi (Otso Ovaskainen), dunson@duke.edu (David B. Dunson).}

\begin{abstract}
Microphone array-based passive acoustic monitoring is increasingly used for biodiversity sensing in forests. However, design and evaluation of array systems and configurations remains difficult since field recordings are costly, difficult to reproduce, and provide limited control over forest and atmospheric conditions. We present \textbf{ForestIR}, a physics-informed and reproducible simulation framework that links forest and environmental conditions to microphone-array recordings for bioacoustic remote sensing. Through a more realistic sound propagation method and a systematic control over array design and environmental factors, ForestIR provides a practical simulation framework for optimizing array-based monitoring systems, especially for sound source localization purposes. ForestIR generates source–microphone impulse responses (IRs) under user-controlled forest and atmospheric conditions, and renders synthetic array recordings by convolving test signals with controlled background noise. 
We evaluate and demonstrate realistic features of ForestIR through experiments based on localization sensitivity to forest layout and atmospheric conditions, and also comparison between simulated IRs with sine-sweep IR measurements from a field experiment. 
ForestIR provides a practical way to test how forest and ground conditions, atmospheric state, and array geometry affect bioacoustic localization, and can support microphone-array design, robustness testing, and synthetic-data generation for passive acoustic monitoring.
\end{abstract}

\vspace{2em}

\newpage

\section{Introduction}
\label{sec:intro}
Passive acoustic monitoring (PAM) has become increasingly important in biodiversity monitoring due to difficulties in detecting wildlife by sight in forests. Compared to traditional surveys, which comes with much higher labor and time cost, PAM can produce data over substantially larger areas and longer periods \citep{PettorelliEtAl2016,AllanEtAl2018Technoecology,BergerTalLahozMonfort2018,Stephenson2020TechAdv,LahozMonfortMagrath2021}. For vocal species such as birds, bats and cicadas, acoustic sensing works especially well, and is now a common tool for biodiversity assessment and conservation \citep{GibbEtAl2019,SugaiEtAl2019,StowellSueur2020}. 

We are particularly motivated by the problem of designing array systems for sound source localization purposes. Simple PAM systems are effective for monitoring the presence or absence of vocal species, but collecting acoustic data with detailed information on sound source locations is a challenging task. Such location information is important because it provides deeper insight into which species are present and where they occur in a detailed manner. By synchronizing the microphone arrays, it is possible to estimate where the calls are originating from based on the time difference of arrival information \citep{Darras2019,Rhinehart2020}. However, sound propagation conditions can vary substantially with vegetation and atmospheric conditions \citep{Rhinehart2020,Verreycken2021HeterogeneousArrays,Haupert2023DetectionDistance,Ostashev2017}, which significantly affect the accuracy of the localization results. These challenges make it difficult to design and evaluate the PAM system using field recording data alone, which is costly, difficult to reproduce, and often multiple factors confounded by many factors at the same time.

Simulation provides a valuable tools to design and evaluate array-based monitoring systems in field recordings because it allows to change array layouts, source positions, forest structure, and weather conditions one at a time. This level of control is valuable for designing arrays, testing localization algorithms, and creating synthetic data for training machine learning algorithms \citep{Rhinehart2020,KanekoGamper2022,Kujawski2024SyntheticDataFramework}. However, current resources only partly meet these needs. Public IR libraries like OpenAIR \citep{Murphy2010} allow auralization and benchmarking, but do not offer flexible models that link forest features to array recordings . Earlier tools, such as the \emph{Forest Reverb Generator} \citep{Wiens2008}, simulate simple trunk scattering but are not built as reproducible, end-to-end pipelines for localization experiments.

Among these acoustic simulators, wave-based methods can capture effects such as diffraction and interference, but are often too computationally heavy for large outdoor areas \citep{Botteldooren1995,RaghuvanshiEtAl2010PrecomputedWave}. Diffuse-field models summarize reverberant energy and late decay, but do not keep the timing needed for classical time difference of arrival (TDoA) and steered response power (SRP)-based localization \citep{Vorlander2008}. By contrast, path-based methods create impulse responses from delayed and weakened sound paths. While this is less complete than full wave models, it works well for generating many clear source–microphone impulse responses when geometry changes \citep{AllenBerkley1979,SaviojaSvensson2015}. In more recent forest acoustics studies, \citet{KanekoGamper2021} showed that single-scattering cylinders can efficiently create forest-like echo patterns, and \citet{KanekoGamper2022} showed their usefulness for bird localization with distributed arrays. Still, these models have limited control over details like exact tree locations, tree height, branch and leaf scattering, ground conditions, and the state of the atmosphere, such as temperature, relative humidity, and air pressure.

Given this background, we introduce ForestIR, a reproducible path-based simulator for array-based bioacoustic remote sensing. ForestIR creates source–microphone impulse responses with adjustable tree geometry, ground effects, atmospheric absorption, and optional branch and leaf scattering. It then makes multichannel recordings by convolving these impulse responses with dry bird calls or test signals and adding controlled background noise. Forest scenes can be built from mapped tree locations or generated synthetically using uniform or repulsive sampling, allowing experiments to change vegetation structure while keeping the scene easy to understand.

We run four experiments to illustrate that ForestIR captures important effects for localization. First, we check if SRP (with phase transform) localization error changes with different vegetation layouts while keeping the source, array, and weather the same. Second, we test if mismatches in atmospheric conditions, especially changes in sound speed due to temperature, affect localization when the system assumes a fixed speed. Third, we compare simulated impulse responses to measured ones from exponential sine sweeps recorded in the snow field in Konnevesi, Finland. Finally, we compare simulated and real bird vocalizations under the same geometry using Mel-Frequency Cepstral Coefficients (MFCC) cosine similarity, spectrogram cross-correlation (SPCC), Acoustic Complexity Index (ACI), and Acoustic Evenness Index (AEI) \citep{DavisMermelstein1980MFCC,SomervuoHarmaFagerlund2006,ClarkMarlerBeeman1987SPCC,Pieretti2011,VillanuevaRivera2011}. These experiments together test if ForestIR matches field-measured decay and is sensitive to environmental and geometric changes that matter for localization.

Because ForestIR lets users change forest structure, ground conditions, atmosphere, array layout, and noise separately, it supports reproducible studies on how these factors affect array recordings and localization. This helps guide decisions about microphone-array design and deployment for biodiversity monitoring.

\section{Materials and Methods}
\label{sec:methods}

\subsection{ForestIR: simulator overview}
\label{subsec:overview}

ForestIR is an open and reproducible simulation framework that connects forest structure, ground and atmospheric conditions, and microphone-array geometry to multichannel bioacoustic recordings for localization research. The simulator implements a lightweight, scriptable pipeline: forest impulse responses (direct path, ground reflection, trunk scattering, and optional canopy scatterers) convolved with dry source waveforms, followed by configurable ambient noise. The pipeline is exposed through a consistent command-line interface and a Python API, both of which support large-scale batch rendering and controlled ablation studies under matched geometry and acoustic settings.

Implementation details, including CLI usage, package entry points, input formats, output manifests, and reproducibility metadata, are provided in Supporting Information.

\subsection{Acoustic Propagation Model}
\label{subsec:propagation_model}

\begin{figure*}[t]
    \centering
    \includegraphics[width=\textwidth]{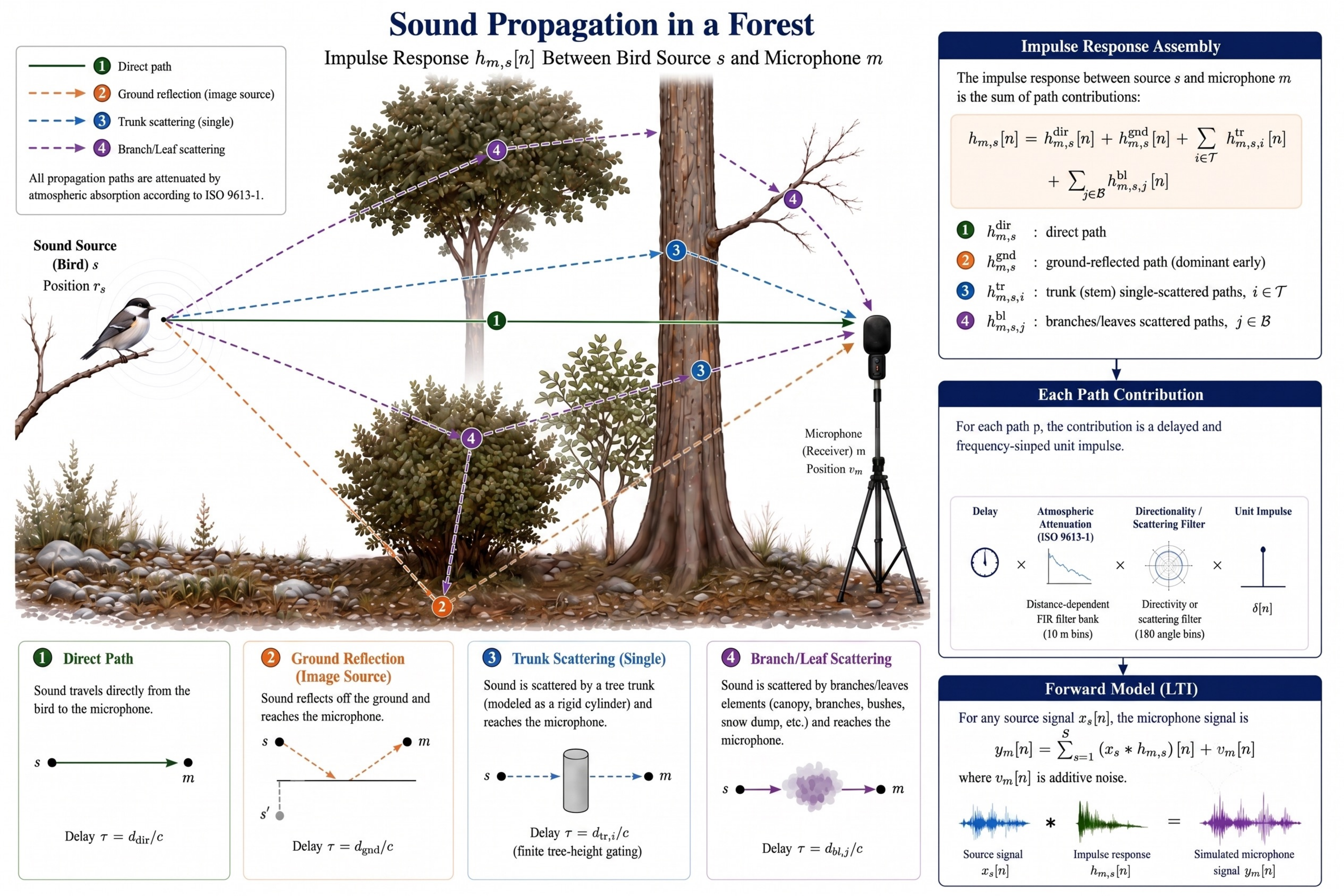}
    \caption{Overview of the ForestIR acoustic propagation model. For each bird source-microphone pair, the simulator assembles an impulse response from delayed and frequency-shaped path contributions, including the direct path, an image-source ground reflection, single scattering from tree trunks, and optional branch and leaf scattering. Each path is attenuated by atmospheric absorption, and the resulting impulse response can be convolved with a dry source signal to generate a simulated microphone recording. The configurable branch and leaf scattering layer can be used to represent aloft canopy scattering, dense undergrowth, shrubs, or other small-scale vegetation structures depending on the simulated scene.}
    \label{fig:forestir_overview}
\end{figure*}

\Cref{fig:forestir_overview} summarizes the propagation components used by ForestIR and illustrates how individual delayed and filtered path contributions are assembled into a source–microphone impulse response.

\subsubsection*{Impulse response and forward model}
ForestIR adopts a linear time-invariant (LTI) forward model between each sound source $s$ and microphone $m$ at sampling rate $f_s$, characterized by a discrete-time impulse response (IR) $h_{m,s}[n]$. For a dry source waveform $x_s[n]$, the simulated microphone signal is
\begin{equation}
y_m[n] \;=\; \sum_{s=1}^{S} (x_s * h_{m,s})[n] \;+\; v_m[n],
\label{eq:lti_forward}
\end{equation}
where $*$ denotes discrete convolution and $v_m[n]$ is an additive noise term (\Cref{subsec:mic_simulation}). This IR-based formulation enables reusing a fixed propagation channel $h_{m,s}[n]$ to render arbitrary source signals under the same geometry and atmospheric state, a standard strategy in acoustic simulation \citep{AllenBerkley1979}.

ForestIR computes $h_{m,s}[n]$ as the simulated response to a unit-sample excitation, i.e., by setting $x_s[n]=\delta[n]$ in \Cref{eq:lti_forward}. The IR is assembled by accumulating a set of path contributions, each implemented as a delayed and frequency-shaped version of the unit impulse. Denoting the source and microphone positions by $\mathbf{r}_s$ and $\mathbf{r}_m$, ForestIR constructs
\begin{equation}
h_{m,s}[n] \;=\; h^{\mathrm{dir}}_{m,s}[n] \;+\; h^{\mathrm{gnd}}_{m,s}[n]
\;+\; \sum_{i\in\mathcal{T}} h^{\mathrm{tr}}_{m,s,i}[n]
\;+\; \sum_{j\in\mathcal{B}} h^{\mathrm{bl}}_{m,s,j}[n],
\label{eq:ir_sum}
\end{equation}
where $h^{\mathrm{dir}}_{m,s}$ is the direct path, $h^{\mathrm{gnd}}_{m,s}$ is the dominant early ground-reflected path (image-source), $\{h^{\mathrm{tr}}_{m,s,i}\}$ are single-scattered contributions from tree stems indexed by $\mathcal{T}$, and $\{h^{\mathrm{bl}}_{m,s,j}\}$ are optional stochastic scattering contributions from branch and leaf elements indexed by $\mathcal{B}$.

Each contribution is synthesized by combining a 3D path length, a propagation delay $\tau=d/c$, frequency-dependent atmospheric attenuation from ISO~9613-1, and, when relevant, a directionality or scattering filter determined by the interaction type \citep{ISO9613-1}. This path-sum representation makes the propagation model spatially explicit while remaining efficient enough for repeated source–microphone simulations. Implementation details, including filter-bank precomputation and batching over scatterers, are provided in Supporting Information.

The trunk-scattering component follows the single-scattering-cylinder (SSC) philosophy used for fast forest reverberation simulation \citep{KanekoGamper2021,KanekoGamper2022}, but ForestIR is designed for localization-oriented experiments with interpretable scene configuration. In particular, (i) all geometric path lengths used for delays and ISO~9613-1 attenuation are computed in full 3D; (ii) scattering elements correspond to either mapped stems (site-specific tree CSVs) or explicitly defined synthetic forests; (iii) finite tree-height gating is applied to trunk scattering; and (iv) scattering from aloft foliage is represented by an explicit branch and leaf layer that can be enabled or disabled independently.

\subsubsection*{Speed of sound model}
Microphone-array localization methods commonly convert measured time differences of arrival into range-difference constraints using an assumed propagation speed \citep{BrandsteinWard2001}. This assumption can become inaccurate outdoors because sound speed varies with atmospheric state, and uncertainty in sound speed can degrade time-difference-based localization \citep{RabensteinAnnibale2017}. ForestIR therefore computes the propagation delay $\tau=d/c$ using a speed of sound $c$ determined by the configured temperature, relative humidity, and air pressure, rather than treating $c$ as a fixed constant. 

We use a lightweight moist-air mixture approximation based on the ideal-gas expression
\begin{equation}
c \;=\; \sqrt{\gamma \, R_{\mathrm{spec}} \, T},
\label{eq:sos_ideal}
\end{equation}
where $T$ is absolute temperature, $R_{\mathrm{spec}}$ is the mixture-specific gas constant, and $\gamma$ is the ratio of specific heats. Following standard thermodynamic treatments of sound speed in moist air \citep{Cramer1993,GaviosoEtAl2025JPR}, ForestIR allows the mixture composition, and hence $R_{\mathrm{spec}}$ and $\gamma$, to vary with humidity and pressure. The detailed calculation is given in Supporting Information. This atmospheric dependence is important for TDoA- and SRP-based localization because even small biases in $c$ can shift the predicted inter-microphone delays over tens of meters.

\subsubsection*{Direct path}
The direct-path distance is $d^{\mathrm{dir}}_{m,s}=\|\mathbf{r}_m-\mathbf{r}_s\|_2$, with delay $\tau^{\mathrm{dir}}_{m,s}=d^{\mathrm{dir}}_{m,s}/c$. The direct contribution is constructed by filtering the unit impulse by the ISO~9613-1 atmospheric attenuation associated with $d^{\mathrm{dir}}_{m,s}$, then applying the corresponding time shift.

\subsubsection*{Ground-reflected path and simplified ground-type parameterization}
ForestIR models the dominant early ground reflection using an image-source construction, with a mirrored source $\mathbf{r}'_s$ reflected across the ground plane. The reflected-path distance is $d^{\mathrm{gnd}}_{m,s}=\|\mathbf{r}_m-\mathbf{r}'_s\|_2$, and the contribution is time-shifted by $d^{\mathrm{gnd}}_{m,s}/c$. The reflected component is scaled by a ground-type-specific amplitude multiplier, denoted by $\alpha_g(\mathrm{type})$, determined by a user-selected ground type. This parameter is used as a simple proxy for the net reflectivity of the ground-reflected path, rather than as a full impedance based reflection coefficient depending on frequency. This parameterization is motivated by outdoor propagation standards and acoustics treatments of ground interaction \citep{ISO9613-2,Moser2009}.

The choice of $\alpha_g(\mathrm{type})$ is guided by standard outdoor acoustics ground classifications and by specific material absorption studies. Following ISO 9613-2, concrete, water, and ice are treated as \emph{hard ground}, i.e.\ low-porosity and strongly reflective surfaces, whereas grass and other vegetated surfaces are treated as \emph{porous ground}. For normal concrete, published reviews report a sound absorption coefficient of approximately 0.02, consistent with a highly reflective surface \citep{FediukEtAl2021}. For grass, measurements have shown that acoustic absorption varies substantially with grass length and soil moisture, supporting the use of a lower effective reflection factor than for hard ground \citep{LondheRaoBlough2009}. Snow is likewise a porous and acoustically absorptive medium, with experimentally measured absorption depending strongly on snow structure and frequency \citep{DattEtAl2016}. We approximate these differences using fixed ground-type-specific amplitude multipliers.We use $\alpha_g(\mathrm{type})$=0.99 for concrete, water, and ice, $\alpha_g(\mathrm{type})$=0.80 for grass, and $\alpha_g(\mathrm{type})$=0.70 for snow. These values should be interpreted as simplified approximations for comparing broad classes of ground conditions, rather than as universal physical reflection coefficients.

\subsubsection*{Trunk scattering}
Trees are represented by mapped stems with per-tree radius $a_i$ and height $H_i$ specified by the user. ForestIR supports both site-specific and synthetic tree layouts. For site-specific simulations, tree stems are read from a user-provided CSV with per-tree position, radius, and height. For synthetic forests, stem locations can be generated either by uniform sampling over the horizontal simulator domain, or by a repulsive point process implemented via rejection sampling, which enforces a user-specified minimum inter-tree spacing. The repulsive option produces more realistic non-overlapping stem configurations and avoids unrealistically clustered trunks, at the cost of occasional rejections when the requested density is high. All stochastic sampling steps are controlled by an explicit random seed recorded in the output manifest.

Trunk scattering is modeled as single scattering from rigid cylinders. For each tree $i$ included in the scattering set, ForestIR constructs a broken path $s \rightarrow i \rightarrow m$ with 3D distance
\begin{equation}
d^{\mathrm{tr}}_{m,s,i} \;=\; \|\mathbf{r}_i-\mathbf{r}_s\|_2 + \|\mathbf{r}_m-\mathbf{r}_i\|_2,
\label{eq:broken_path}
\end{equation}
and delay $\tau^{\mathrm{tr}}_{m,s,i}=d^{\mathrm{tr}}_{m,s,i}/c$. 

To represent variability in trunk size while keeping simulations efficient, ForestIR supports per-tree trunk diameters either provided directly in the input stem map or sampled from a user-specified diameter range. Because rigid-cylinder scattering filters depend on trunk radius, we discretize diameters into a small number of radius bins and precompute one directional scattering filter bank per bin; each tree is assigned to the nearest bin and uses the corresponding precomputed filter. Each trunk contribution is then attenuated by geometric spreading and ISO~9613-1 air absorption along the broken path.

To handle finite tree height in 3D configurations, ForestIR applies a geometric gating rule: trunk scattering from tree $i$ is included only when both the source and microphone heights lie below the tree height, i.e., $z_s < H_i$ and $z_m < H_i$. When this condition is not satisfied, the trunk contribution from that tree is omitted.

\subsubsection*{Optional branch and leaf scattering}
ForestIR includes an optional scattering layer to represent stochastic interactions with aloft branches and leaves. When enabled, we generate a set of scatterers with 3D positions sampled uniformly over the horizontal simulator domain $(x,y)\in[\mathrm{x}_{\min},\mathrm{x}_{\max}]\times[\mathrm{y}_{\min},\mathrm{y}_{\max}]$ and within a user-specified vertical band $z\in[z_{\min},z_{\max}]$. Each scatterer $j$ induces a broken path $s\rightarrow j \rightarrow m$ with 3D path length $d^{\mathrm{bl}}_{m,s,j}=\|\mathbf{r}_j-\mathbf{r}_s\|_2+\|\mathbf{r}_m-\mathbf{r}_j\|_2$ and corresponding delay $d^{\mathrm{bl}}_{m,s,j}/c$.

We model each branch and leaf element as an approximately omnidirectional secondary scatterer with a controllable amplitude parameter, implemented as an impulse-shaped scattering filter. This component can be toggled independently.

\subsection{Microphone Array Recording Simulation}
\label{subsec:mic_simulation}

Given source signals $\{x_s[n]\}_{s=1}^{S}$ and simulated impulse responses $\{h_{m,s}[n]\}$, ForestIR renders clean microphone-array recordings by convolution,
\begin{equation}
\tilde{y}_m[n] \;=\; \sum_{s=1}^{S} (x_s * h_{m,s})[n],
\label{eq:render_clean}
\end{equation}
where $m$ indexes microphones. Input waveforms are resampled to the simulation sampling rate when needed so that time delays and frequency-dependent attenuation are applied consistently. Additive background noise can then be included as
\begin{equation}
y_m[n] \;=\; \tilde{y}_m[n] + v_m[n].
\label{eq:add_noise}
\end{equation}
ForestIR supports both synthetic noise models and recorded environmental noise, with a user-controlled noise level. Details of the available noise types, channel normalization, and recorded-noise sampling procedure are provided in Supporting Information.

\subsection{Acoustic Simulator Validation Data and Metrics}
\label{subsec:prop_eval}

We validate ForestIR at two complementary levels: (i) \emph{impulse-response (IR) fidelity} under matched source--receiver geometries, and (ii) \emph{rendered bird-audio similarity} when dry bird calls are propagated through the simulated forest scene. Throughout, we compare simulated outputs against field recordings from the Konnevesi snow-field playback experiment. Separately, the Konginkangas Trotting Center site provides the measured tree-position map shown in \Cref{fig:konginkangas_tree_detect_clicks}, which is used in the localization sensitivity experiment.

\subsubsection*{Konnevesi snow-field playback recordings}
\label{subsec:data}
Validation data were obtained from controlled playback experiments conducted on a frozen lake at the Konnevesi Research Station, Finland, in February 2026. Audio playback included sine sweeps and bird vocalizations emitted from a known source location, with and without background audio. Recordings were acquired using a three-microphone array with an approximately equilateral geometry and inter-microphone spacing of approximately 50 m. Audio was played from five speaker positions within and slightly outside of the array triangle. All microphones and speakers were positioned at a height of 1.2 m.

Raw audio streams were synchronized across devices to an accuracy corresponding to approximately 1 cm in acoustic propagation distance using PPS (pulse-per-second) timing signals. The synchronized recordings were subsequently segmented into clips ranging from 2.18 to 11.00 s in duration (mean: 5.07 s), each containing a single target signal: either a sine sweep or bird vocalization.

Hourly weather covariates, including temperature, relative humidity, pressure, snow depth, and wind speed, were obtained from Open-Meteo.

\subsubsection*{Measured impulse responses from exponential sine sweeps (ESS)}

Measured IRs were recovered from exponential sine sweep (ESS) recordings using inverse filtering following \citet{Farina2000ESS}. Each recovered per-channel IR provides a measured source--microphone transfer function that can be compared directly with the simulated IR generated under the matched geometry.

\subsubsection*{IR-level comparison via energy decay curves (EDC)}
To assess whether ForestIR reproduces the temporal decay characteristics of the forest, we compute energy decay curves (EDCs) from both measured and simulated IRs using Schroeder backward integration \citep{Schroeder1965}. For an IR $h[n]$, the (unnormalized) EDC is
\begin{equation}
E[n] \;=\; \sum_{k=n}^{N-1} h^2[k],
\label{eq:edc}
\end{equation}
and we normalize $E[n]$ to unit value at $n=0$ for comparability across channels and conditions. Similarity between measured and simulated decay profiles is summarized by the Pearson correlation between their normalized EDCs, computed per microphone channel and then aggregated across channels.

\subsubsection*{Rendered bird-audio similarity}
To evaluate whether ForestIR preserves perceptual and ecological characteristics of bird calls after propagation, we use audio of $30$ dry bird sounds from the BirdNet-derived dataset (duration $2.55$~s) and compare simulated and measured array recordings under matched geometries. For each clip, we compute a set of complementary similarity measures between the measured and simulated waveforms, and report aggregated statistics across all microphone geometries:
(i) \textbf{MFCC cosine similarity}, computed from Mel-frequency cepstral coefficients (MFCCs) to capture perceived timbre and spectral envelope \citep{DavisMermelstein1980MFCC,SomervuoHarmaFagerlund2006};
(ii) \textbf{spectrogram cross-correlation (SPCC)}, which measures similarity in time--frequency structure between the two signals \citep{ClarkMarlerBeeman1987SPCC};
(iii) \textbf{Acoustic Complexity Index (ACI)} \citep{Pieretti2011}; and
(iv) \textbf{Acoustic Evenness Index (AEI)} \citep{VillanuevaRivera2011}.
These metrics jointly quantify perceptual similarity (MFCC, SPCC) and soundscape-style structure (ACI, AEI), providing a practical evaluation of the simulator for downstream bioacoustic monitoring applications.

\subsection*{Use of AI}
Gemini was used only to assist with the preparation and visual refinement of the illustrative elements in \Cref{fig:forestir_overview}, a non-data schematic overview figure.

\section{Results}
\label{sec:results}

\subsection{Localization sensitivity to tree spatial layout}
\label{subsec:results_tree_layout}

\begin{figure}[h]
    \centering
    \includegraphics[width=0.5\linewidth]{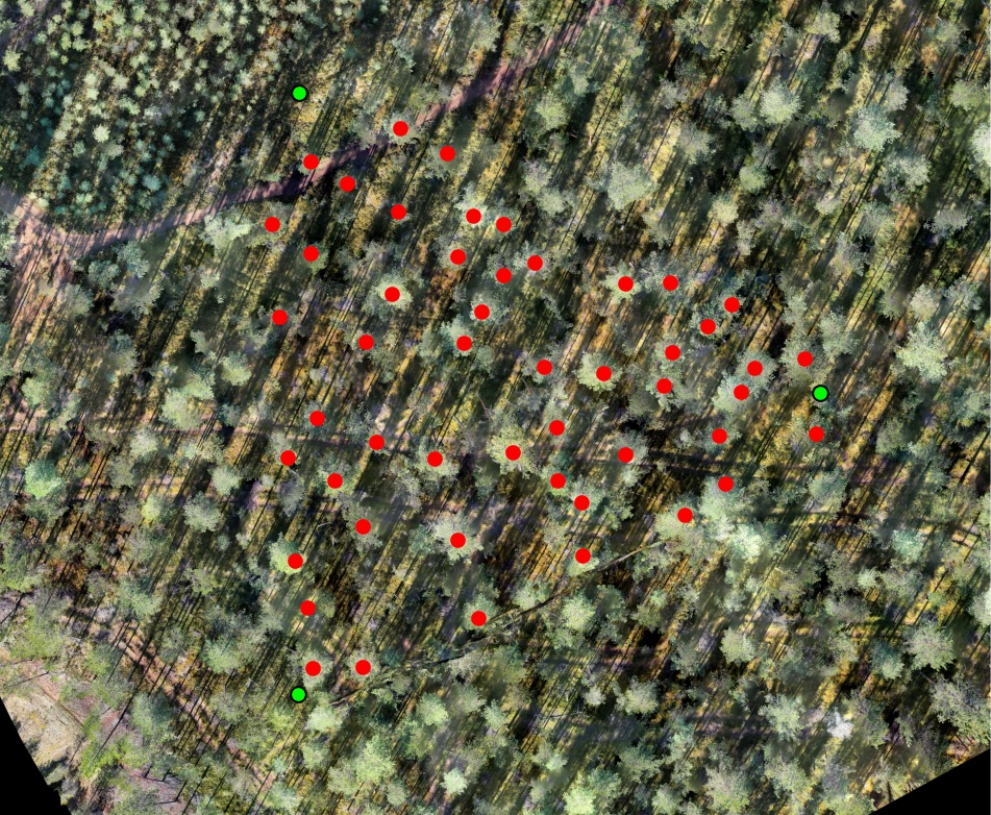}
    \caption{Konginkangas tree-position map used as the measured-tree-layout input in Scenario~1. Red dots indicate manually identified tree locations from the LiDAR-derived site map, and green dots represent the microphone geometry used for the corresponding simulation setup.}
    \label{fig:konginkangas_tree_detect_clicks}
\end{figure}

To investigate whether vegetation geometry alone can influence acoustic localization performance under identical atmospheric conditions, we conducted a controlled simulation study using ForestIR and compared the results against the legacy simulator of \citet{KanekoGamper2021}. One scenario used a measured Konginkangas tree-position map as structural tree geometry input to the simulator, as shown in \Cref{fig:konginkangas_tree_detect_clicks}.

Unless otherwise stated, localization experiments used the same setup: $T=20^\circ$C, RH$=50\%$, $p=1013.25$~hPa, sampling rate $f_s=384$~kHz, no additive noise, and the BirdNET clip \texttt{BirdNET\_01\_XC169082.wav} (\emph{American Goldfinch}, \emph{Spinus tristis}). Localization used three microphones (F3, F4, and F12) from the Konnevesi playback array at height $z=1.2$~m. Source locations formed an $11\times11$ grid with $x\in\{-10,0,\dots,90\}$, $y\in\{0,-10,\dots,-100\}$, and source height $z=1.5$~m, yielding 121 positions. All runs used SRP-PHAT with fixed assumed sound speed $343$~m/s, search-grid resolution $0.05$~m, and 2D Euclidean localization error as the performance metric. For ForestIR tree-layout experiments, branch and leaf scattering used 5000 scatterers sampled over $z\in[2,5]$~m, all trunks had diameter $d_{\mathrm{trunk}}=0.5$~m, and stochastic components used fixed random seeds for reproducibility.

We evaluated five tree-layout scenarios:
\begin{enumerate}
    \item the measured Konginkangas tree map;
    \item a synthetic repulsive-layout forest with 43 trees;
    \item an extreme ``crowded'' configuration in which all 43 trees were artificially collapsed onto the microphone location F4 at $(0,0)$;
    \item the legacy simulator of \citet{KanekoGamper2021} with 43 trees;
    \item the same legacy simulator with 100000 trees distributed over the same forest bounds.
\end{enumerate}
The legacy simulator does not support explicit tree-coordinate specification or branch and leaf scattering, whereas ForestIR allows direct control over tree locations and scattering geometry.

\Cref{tab:tree_layout_results} summarizes how localization performance changes when only the vegetation-scatterer layout is varied.

\begin{table}[h]
\centering
\setlength{\tabcolsep}{4pt}
\resizebox{\textwidth}{!}{
\begin{tabular}{lccccc}
\toprule
Scenario & Tree layout & Branch leaf scattering & Mean error (m) & Median error (m) & Max error (m) \\
\midrule
ForestIR: Konginkangas layout & Konginkangas tree map & Yes & 0.8141 & 0.2236 & 17.9790 \\
\addlinespace
ForestIR: Repulsive layout & Repulsive process (43 trees) & Yes & 0.8145 & 0.2236 & 17.9790 \\
\addlinespace
ForestIR: Crowded layout & 43 trees collapsed at F4 & Yes & 97.5471 & 100.4988 & 162.7882 \\
\addlinespace
Legacy model: 43 trees & Implicit/random forest & No & 0.1762 & $\approx 0$ & 15.7590 \\
\addlinespace
Legacy model: 100000 trees & Implicit/random forest & No & 0.1493 & $\approx 0$ & 16.2117 \\
\bottomrule
\end{tabular}
}
\caption{Localization performance under identical atmospheric conditions but different tree spatial layouts. The two ForestIR layouts produced similar errors, and collapsing 43 trees at microphone F4 caused a large localization failure as expected in real forest. In contrast, the legacy model showed limited sensitivity to increasing tree count. These results illustrate why localization experiments benefit from a physically interpretable simulator with explicit vegetation geometry: tree placement, not only tree count, can substantially alter the propagation field and localization outcome.}
\label{tab:tree_layout_results}
\end{table}

The two realistic ForestIR layouts (measured Finland map and repulsive synthetic forest) produced similar localization behavior, with mean localization error approximately $0.81$~m. When 43 trees were artificially concentrated near one microphone, localization performance degraded substantially, with mean error increasing to $97.5$~m.

These results demonstrate that localization behavior can depend strongly on vegetation geometry. This is consistent with the interpretation that strong, spatially concentrated scattering near one microphone can distort inter-channel timing cues and thereby degrade the SRP-PHAT objective.

By contrast, the legacy simulator showed comparatively limited sensitivity. Even when the number of trees increased to $100000$, the mean localization error remained below $0.15$~m. This shows that simplified forest scattering models without explicit spatial control may underestimate localization degradation induced by complex vegetation structure.

Overall, these experiments support the central motivation of ForestIR: physically interpretable and spatially controllable forest acoustic simulation is important for evaluating localization robustness under realistic environmental variability.

\subsection{Localization sensitivity to atmospheric temperature}
\label{subsec:results_temperature_sensitivity}

To test whether atmospheric state alone can influence localization performance, we repeated the same SRP-PHAT localization setup described above while varying only the simulated atmospheric temperature. The tree layout, microphone geometry, 121-position source grid, BirdNET vocalization, rendering pipeline, and localization settings were held fixed. ForestIR computed the propagation speed $c_{\mathrm{sim}}$ from the atmospheric state, whereas SRP-PHAT used the same fixed assumed speed $343$~m/s. This isolates the effect of sound-speed mismatch between the simulated acoustic propagation and the localization model. We additionally evaluated one fixed source position, $(14.94,-64.23)$, using 30 BirdNET vocalizations to test sensitivity to acoustic content.

\begin{table}[h]
\centering
\setlength{\tabcolsep}{4pt}
\resizebox{\textwidth}{!}{
\begin{tabular}{lcccccc}
\toprule
\multicolumn{7}{c}{\textbf{Panel A: 121 source positions, single BirdNET vocalization}} \\
\midrule
Temperature ($^\circ$C) & $c_{\mathrm{sim}}$ (m/s) & Trials & Mean error (m) & Median error (m) & Std. deviation (m) & Max error (m) \\
\midrule
0  & 331.45 & 121 & 4.763 & 2.602 & 8.075 & 53.652 \\
5  & 334.54 & 121 & 2.507 & 1.651 & 2.329 & 10.470 \\
10 & 337.63 & 121 & 1.763 & 0.985 & 2.191 & 15.963 \\
15 & 340.72 & 121 & 0.947 & 0.500 & 1.476 & 9.216 \\
20 & 343.84 & 121 & \textbf{0.739} & \textbf{0.224} & 2.019 & 17.979 \\
25 & 346.99 & 121 & 1.666 & 0.791 & 4.373 & 46.345 \\
30 & 350.18 & 121 & 3.556 & 1.412 & 10.916 & 107.255 \\
\midrule
\multicolumn{7}{c}{\textbf{Panel B: source position (14.9, -64.2), 30 BirdNET vocalizations}} \\
\midrule
Temperature ($^\circ$C) & $c_{\mathrm{sim}}$ (m/s) & Mean estimate $(\hat{x},\hat{y})$ & Mean error (m) & Std. deviation (m) & \multicolumn{2}{c}{} \\
\midrule
0  & 331.45 & $(13.75,\,-65.87)$ & 2.33 & 8.4 & \multicolumn{2}{c}{} \\
5  & 334.54 & $(21.10,\,-55.76)$ & 26.0 & 23.4 & \multicolumn{2}{c}{} \\
10 & 337.63 & $(16.44,\,-62.17)$ & 8.14 & 22.0 & \multicolumn{2}{c}{} \\
15 & 340.72 & $(14.83,\,-64.38)$ & 0.48 & 1.2 & \multicolumn{2}{c}{} \\
20 & 343.84 & $(14.86,\,-64.34)$ & \textbf{0.14} & 0.03 & \multicolumn{2}{c}{} \\
25 & 346.99 & $(15.00,\,-64.05)$ & 0.19 & 0.0 & \multicolumn{2}{c}{} \\
30 & 350.18 & $(15.10,\,-64.00)$ & 0.28 & 0.0 & \multicolumn{2}{c}{} \\
\bottomrule
\end{tabular}
}
\caption{Localization sensitivity to atmospheric temperature. ForestIR used the temperature-dependent sound speed $c_{\mathrm{sim}}$, while SRP-PHAT used a fixed $343$~m/s. Panel A reports 121 source positions for one BirdNET call; Panel B reports 30 BirdNET calls at one fixed source position. Errors were smallest near the matched sound-speed condition and increased under stronger temperature mismatch, highlighting the importance of simulators that expose atmospheric temperature as a configurable input rather than assuming a fixed sound speed.}
\label{tab:temperature_sensitivity}
\end{table}

Localization error was smallest near $20^\circ$C, where $c_{\mathrm{sim}}$ was closest to the sound speed assumed by SRP-PHAT. Error increased as temperature moved away from this matched condition, with maximum errors reaching 53.7~m at $0^\circ$C and 107.3~m at $30^\circ$C. The 30-call experiment showed the same pattern but also revealed call-dependent failures: at $5^\circ$C, mean error increased to 26.0~m with standard deviation 23.4~m. Thus, atmospheric mismatch can interact with signal content, causing severe SRP-PHAT errors for some vocalizations.

Although the localization experiment above varies temperature, ForestIR also allows relative humidity and air pressure to affect sound speed. At $T=20^\circ\mathrm{C}$ and $p=101.325$kPa, the legacy temperature-only approximation gives a constant value of $c$, whereas the updated moist-air calculation increases with relative humidity. The detailed calculation and example validation values are provided in Supporting Information.

Together, these experiments demonstrate that a simulator that exposes temperature, humidity, and pressure as controllable parameters is necessary for evaluating when localization algorithms are robust to atmospheric variation and when they require compensation or recalibration.

\subsection{Propagation fidelity against field measurements}
\label{subsec:results_prop}

We then evaluated whether ForestIR reproduces the broad decay characteristics of measured impulse responses under matched source--receiver geometries. Measured IRs were recovered from ESS recordings collected in the February 2026 Finland field experiment, and were compared against simulated IRs generated with the corresponding scene geometry. The simulated scene was configured to approximate the winter snow-field conditions: ForestIR used branch and leaf scattering module with 5000 near-surface scatterers over the height range $z\in[0,0.5]$~m to approximate snow piles and small-scale snow-surface roughness, and used the recorded atmospheric state ($T=-12.8^\circ$C, RH$=32\%$, $p=1013.25$~hPa) together with the snow ground type. No additional background noise was added to the simulated IRs. For comparison, we also considered a baseline diffuse-scattering configuration with 1000 randomly sampled near-surface scatterers over the same height range, intended as a simpler approximation to unmodeled terrain and surface irregularities, and the legacy trunk-scattering model of \citet{KanekoGamper2021}.

\begin{table}[h]
\centering
\setlength{\tabcolsep}{4pt}
\resizebox{\textwidth}{!}{
\begin{tabular}{lcccc}
\toprule
Method & Number of pairs & Mean EDC Pearson & Median EDC Pearson & EDC Pearson range \\
\midrule
ForestIR & 7 & \textbf{0.837} & \textbf{0.843} & 0.774--0.871 \\
Legacy model & 7 & 0.251 & 0.221 & 0.178--0.466 \\
\bottomrule
\end{tabular}
}
\caption{Matched-pair EDC agreement between measured IRs and simulated IRs for seven source--receiver geometries. Pearson correlation was computed between normalized EDCs. ForestIR produced consistently higher EDC agreement than the legacy model, indicating that the explicit near-surface scattering configuration better reproduced the measured decay structure in this winter field setting.}
\label{tab:ir_validation_summary}
\end{table}

\subsubsection*{Measured vs simulated IRs under matched geometry}

\begin{figure}[h]
    \centering
    \includegraphics[width=0.6\linewidth]{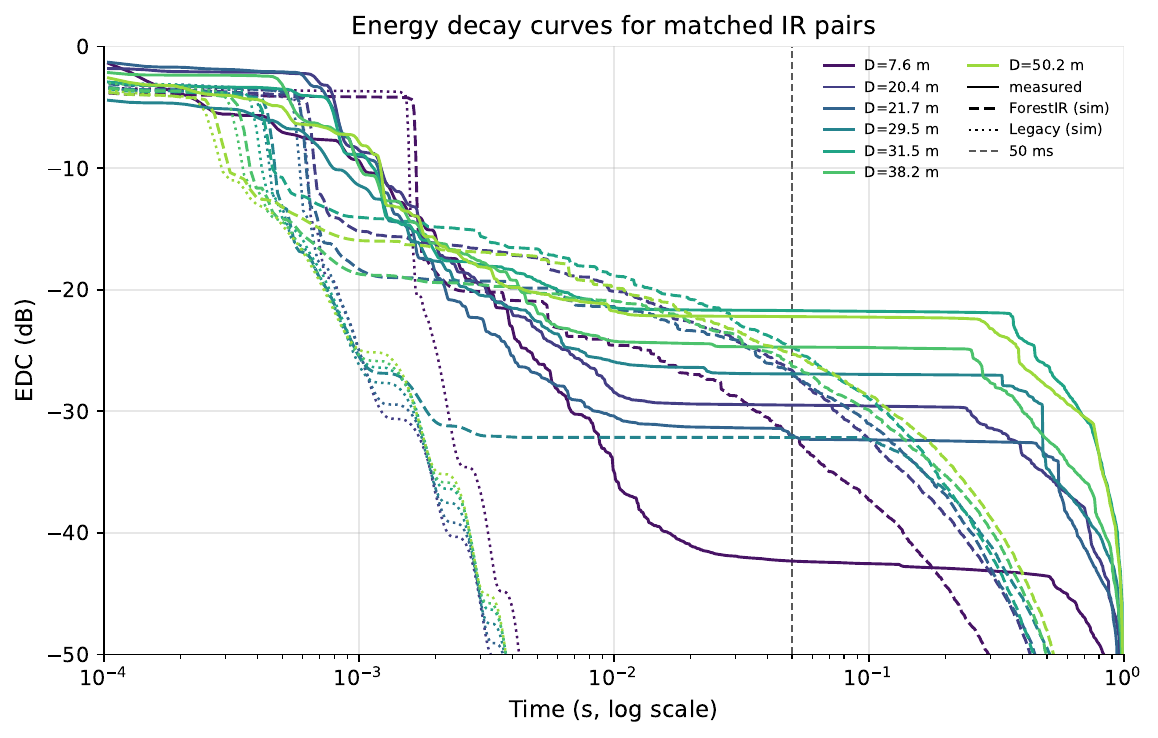}
    \caption{Normalized energy decay curves (EDCs) for seven matched measured and simulated IR pairs spanning source--receiver distances from 7.6~m to 50.2~m. ForestIR follows the measured decay profiles more closely across the matched geometries, whereas the legacy model exhibits substantially faster decay and fails to reproduce the measured long-tail energy. Energy decay curves show good agreement in overall shape and reverberation time, though the simulator overestimates early-time energy relative to field measurements, likely reflecting the anomalous absorption properties of the snow-covered ice surface.}
    \label{fig:edc_matched_pairs}
\end{figure}

\Cref{fig:edc_matched_pairs} compares normalized EDCs for seven matched measured--simulated IR pairs spanning source--receiver distances from 7.6~m to 50.2~m. The time axis is shown on a logarithmic scale to make both the early arrival region and the late decay tail visible, and the vertical 50ms reference line marks the boundary used in the clarity measure $C_{50}$. Across these matched geometries, ForestIR reproduces the overall decay structure of the measured IRs, including both the early decay and the long-tail behavior. This visual agreement is supported by the matched-pair EDC correlations in \Cref{tab:ir_validation_summary}, where ForestIR achieved a mean Pearson correlation of 0.837 and a median correlation of 0.843.

By contrast, the legacy model exhibited substantially faster EDC decay and much weaker agreement with the measured IRs, with mean Pearson correlation of 0.251 and median correlation of 0.221. In this winter snow-field setting, the legacy decay curves drop too rapidly to match the measured long-tail behaviour. A likely reason is that the legacy model is centered on trunk-based scattering and does not explicitly represent near-surface snow piles, small-scale snow roughness, or snow-ground conditions in the same configurable way as ForestIR.

Some discrepancy between measured and simulated IRs remains even for ForestIR, which is expected for an efficient geometric path-based simulator. ForestIR represents reflections and scattering from explicitly modeled objects or scatterer populations, but real outdoor environments contain many sources of diffuse scattering, including fine-scale snow roughness, small branches, terrain irregularities, and device response effects. There may also have been atmospheric layering or local microclimatic effects affecting propagation, potentially influenced by nearby running water. These unmodeled effects mainly influence the detailed late-reverberation tail. For localization-oriented applications, the direct path and early arrivals carry the primary timing information used by TDoA- and SRP-based methods, making exact reproduction of all late diffuse scattering less critical than preserving broad decay behaviour and early-arrival structure.

\subsubsection*{Baseline comparison and distance dependence}

\begin{figure}[h]
    \centering
    \includegraphics[width=0.6\linewidth]{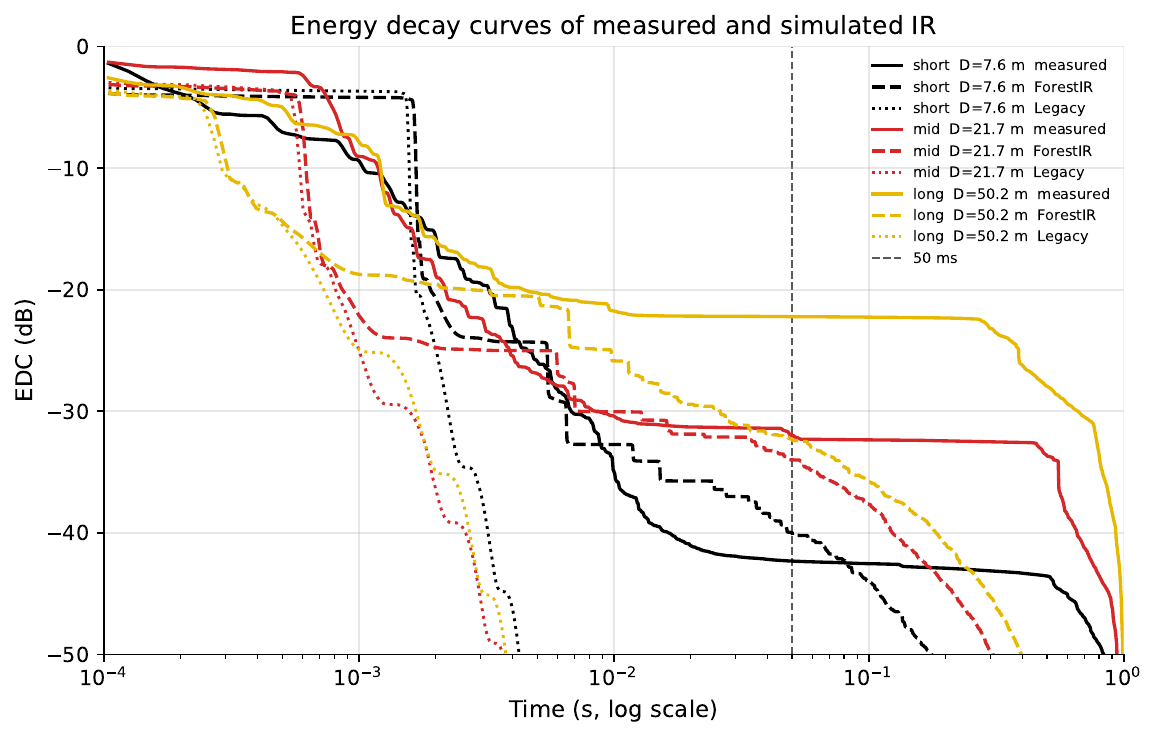}
    \caption{Baseline comparison of normalized EDCs at three representative source–receiver distances. ForestIR better follows the measured distance-dependent decay behaviour, whereas the legacy model exhibits substantially faster decay and underestimates the measured long-tail energy. The time axis is logarithmic, and the vertical 50~ms line marks the early/late energy boundary used in $C_{50}$.}
    \label{fig:edc_baseline_compare}
\end{figure}

To compare the decay behavior across representative propagation distances, we selected three matched geometries corresponding to short ($D=7.6$~m), intermediate ($D=21.7$~m) and long ($D=50.2$m) source–receiver separations. \Cref{fig:edc_baseline_compare} compares the measured EDCs with two simulated configurations: the baseline diffuse-scattering configuration with 1000 randomly sampled scatterers over $z\in[0,0.5]$~m, and the legacy trunk-scattering model. The baseline diffuse-scattering configuration is included as a simple reference for unresolved environmental terrain effects.

The measured EDCs decay more slowly as source–receiver distance increases, indicating that late-arriving energy becomes relatively more prominent at larger separations. The baseline diffuse-scattering ForestIR model partially increases late energy and remains close to the measured EDCs. By contrast, the legacy model decays substantially faster across the representative distances, consistent with its lack of an explicit near-surface diffuse-scattering mechanism for this snow-field setting.

\subsubsection*{Scatterer-count sensitivity}
\begin{figure}[h]
\centering
\includegraphics[width=0.6\linewidth]{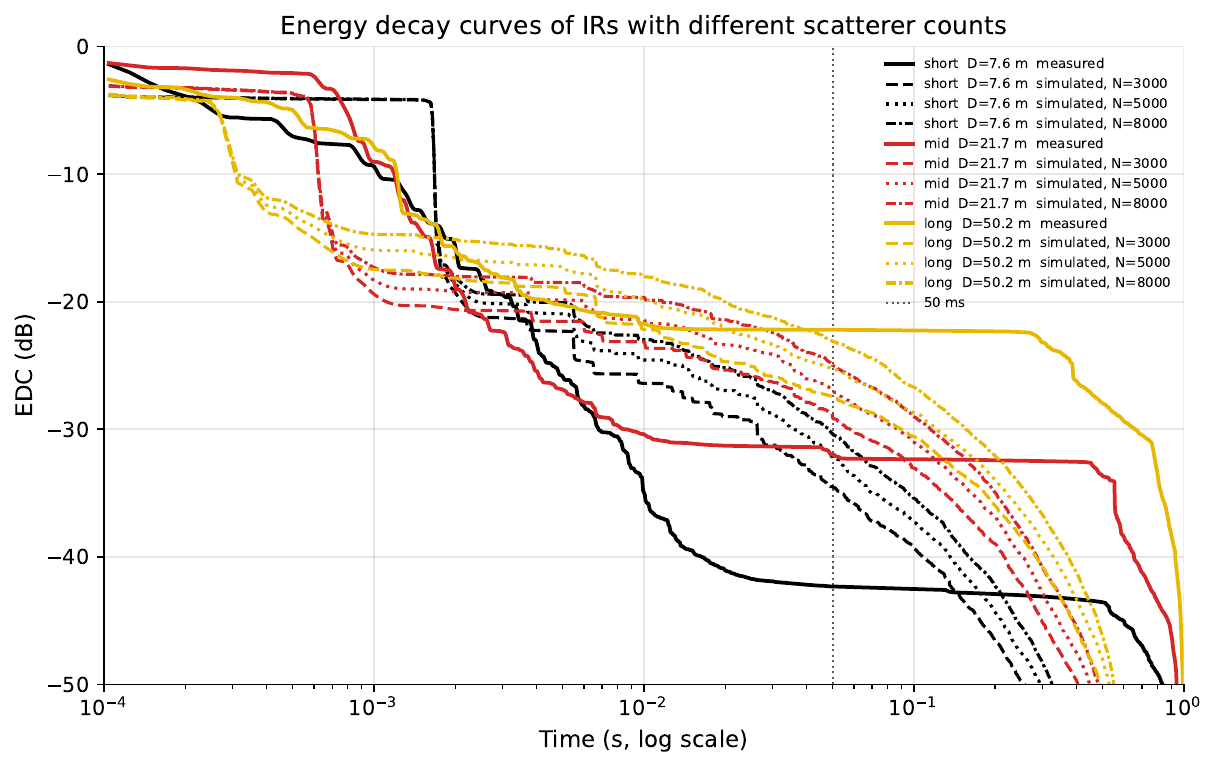}
\caption{Sensitivity of ForestIR EDCs to the number of near-surface scatterers used to approximate snow piles and small-scale snow-surface roughness. Simulations are shown for 3000, 5000, and 8000 scatterers at short, intermediate, and long source–receiver distances. Increasing the number of scatterers generally slows the simulated energy decay, reflecting stronger late-energy contributions from near-surface scattering. The time axis is logarithmic, and the vertical 50~ms line marks the early/late energy boundary used in $C_{50}$.}
\label{fig:edc_scatterer_count}
\end{figure}

We further examined how the number of near-surface scatterers affects the simulated decay behaviour in ForestIR. \Cref{fig:edc_scatterer_count} compares ForestIR simulations using 3000, 5000, and 8000 scatterers at the same three representative distances. Increasing the number of scatterers generally slows the simulated energy decay, consistent with the interpretation that additional near-surface scattering increases the relative contribution of later-arriving energy. This sensitivity analysis illustrates how the branch and leaf scattering module, repurposed here as near-surface scatterers for snow piles, provides a tunable mechanism for matching broad measured decay behaviour while retaining explicit control over the simulated scene.

\subsubsection*{Early-to-late energy balance}

\begin{figure}[h]
    \centering
    \includegraphics[width=0.6\linewidth]{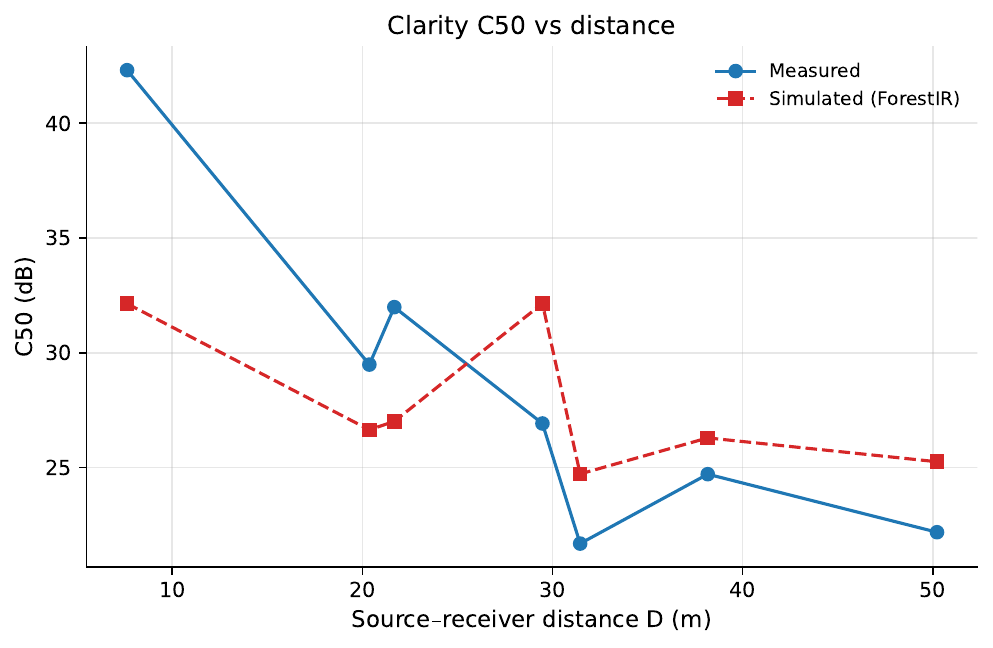}
    \caption{Clarity $C_{50}$ as a function of source--receiver distance for measured and ForestIR-simulated IRs. $C_{50}$ quantifies the ratio of early energy arriving within the first 50~ms to later energy arriving after 50~ms. Both measured and simulated IRs show decreasing $C_{50}$ with increasing distance, indicating that late energy becomes relatively more prominent at larger source--receiver separations. ForestIR captures this directional trend, although a systematic offset remains.}
    \label{fig:c50_vs_distance}
\end{figure}

The clarity measure $C_{50}$ provides a scalar summary of early versus late energy (\Cref{fig:c50_vs_distance}). Both measured and simulated IRs show decreasing $C_{50}$ with distance, indicating that late energy becomes relatively more prominent at larger source--receiver separations. ForestIR captures this directional trend, although a systematic offset remains. Together, the EDC and $C_{50}$ results suggest that ForestIR captures both matched-pair decay behaviour, and the expected directional trend of distance-dependent reverberation structure.

\clearpage
\subsection{Validation with field bird calls data}
\label{subsec:results_audio}

Finally, we compared ForestIR-rendered bird-call snippets with field recordings from the snow-covered open site. Because the propagation region contained no trees, both ForestIR and the legacy model were run with zero trees; ForestIR used snow ground and Konnevesi background noise with \texttt{noise\_level}=0.3. We evaluated three microphones, six source positions, and 30 BirdNET clips, yielding $540$ simulated--recorded comparisons per method. Each pair was aligned by cross-correlation and truncated to a common duration before computing MFCC cosine similarity, SPCC, ACI absolute difference, and AEI absolute difference.

\begin{table}[h]
\centering
\setlength{\tabcolsep}{4pt}
\resizebox{\textwidth}{!}{
\begin{tabular}{lcccc}
\toprule
Method & MFCC cosine $\uparrow$ & SPCC $\uparrow$ & ACI abs. diff. $\downarrow$ & AEI abs. diff. $\downarrow$ \\
\midrule
Dry source & 0.94090 & 0.42512 & 0.10190 & \textbf{0.06799} \\
Kaneko et al. baseline & 0.95691 & 0.26942 & 0.09599 & 0.07272 \\
ForestIR & \textbf{0.98377} & \textbf{0.47845} & \textbf{0.06686} & 0.07284 \\
\bottomrule
\end{tabular}
}
\caption{Mean similarity between simulated and field-recorded bird-call snippets across 540 comparisons. The comparison used three microphones, six source positions, and 30 BirdNET vocalizations under the same alignment and metric pipeline. Higher values indicate better agreement for MFCC cosine similarity and SPCC, whereas lower values indicate better agreement for ACI and AEI absolute differences. ForestIR achieved the best agreement on three of the four metrics, suggesting improved reproduction of spectral envelope, time--frequency structure, and short-timescale acoustic variability relative to the dry-source and legacy-model baselines.}
\label{tab:field_similarity}
\end{table}

ForestIR achieved the best agreement on three of the four metrics, with the highest MFCC cosine similarity and SPCC and the lowest ACI absolute difference (\Cref{tab:field_similarity}). These indicate closer agreement in average spectral envelope, time--frequency structure, and short-timescale acoustic variability. AEI was similar across methods and was slightly best for the dry-source baseline, suggesting that broad spectral evenness in this setting may be driven more by the original bird-call spectrum than by propagation.

\section{Discussion}
\label{sec:discussion}
ForestIR was developed to make forest sound simulation useful for microphone array localization studies. In the validation experiments, ForestIR reproduced the broad decay behavior of measured impulse responses more closely than the legacy trunk-scattering model. It also produced rendered bird call snippets that matched field recordings more closely in MFCC similarity, SPCC, and ACI. These results suggest that the model captures propagation effects visible in both IR decay and propagated bioacoustic signals.

The localization experiments show why this control is useful. When the array geometry, source grid, and localization method were held fixed, changing either the vegetation layout or the atmospheric temperature changed SRP-PHAT localization error. The tree-layout experiment shows that scatterer placement can matter as much as the number of scatterers, as concentrating trees near a single microphone strongly distorted the localization results. The temperature experiment shows that a mismatch in propagation speed can largely affect localization when the localizer assumes a fixed sound speed. These results show that localization studies can miss important failure modes if they use simulators with fixed atmosphere assumptions or only coarse control over forest geometry.

ForestIR still makes simplifying assumptions. Trunk scattering is modeled with first-order scattering only. The model does not fully capture multiple scattering or fine-scale diffuse scattering from snow and small vegetation. The ground model is also simplified by using amplitude multipliers specific to ground types, rather than a full impedance model that accounts for frequency. However, these approximations mainly affect detailed late reverberation. For localization and array-design studies, the direct path and early arrivals carry the dominant cues, so exact reproduction of every late diffuse component is less critical than preserving the main delay structure and broad decay behavior.

Overall, ForestIR provides a useful tool for controlled experiments in array-based bioacoustic monitoring. By allowing vegetation geometry, ground condition, atmospheric state, array layout, and noise to be varied independently, it supports microphone-array design, localization stress testing, and synthetic training-data generation for biodiversity monitoring applications.


\section*{Acknowledgments}
The authors thank Patrik Lauha for his helpful discussions. We also thank Ala-Keiteleen Hevosystäväin seura and the Konginkangas Trotting Center for granting permission to conduct the field experiment at their premises. This work was supported in part by the U.S. National Science Foundation under Grant No. 2426762, the European Research Council (ERC) under the European Union’s Horizon 2020 research and innovation programme (Grant agreement No. 856506), the Research Council of Finland under the Grants No. 336212, 345110 and 367674, and by the Jane and Aatos Erkko Foundation.

\section*{Declaration on the Use of AI}
Gemini was used only to assist with the preparation and visual refinement of the illustrative elements in \Cref{fig:forestir_overview}, a non-data schematic overview figure. No AI tools were used to perform data analysis, alter data visualizations, write manuscript text, or generate scientific conclusions.

\section*{Author contributions}
Conceptualization, X.S.; Methodology, X.S., J.K., C.L., and D.B.D.; Software, X.S.; Validation, X.S.; Formal analysis, X.S.; Investigation, X.S.; Data curation, X.S., J.K., C.L., B.S, P.S, A.L., S.vB.; Writing—original draft preparation, X.S.; Writing—review and editing, X.S., C.L., B.S., J.K., O.O. and D.B.D.; Supervision, D.B.D.; Funding acquisition, D.B.D., O.O.; Field data design and implementation by A.L., O.N., P.S., S.vB., O.O. All authors have read and agreed to the published version of the manuscript.

\section*{Data and Code Availability}
\label{sec:availability}
Code for the ForestIR simulator is available at GitHub {\hypersetup{urlcolor=magenta}\url{https://github.com/TIPColin/ForestIR}}. Processed site-recorded data necessary to reproduce the main analyses can be requested from authors.

\section*{Conflict of interest}
The authors declare no potential conflict of interests.

\bibliographystyle{plainnat}
\bibliography{refs}

\newpage
\clearpage

\setcounter{section}{0}
\setcounter{equation}{0}
\setcounter{table}{0}

\renewcommand{\theequation}{S\arabic{equation}}
\renewcommand{\thetable}{S\arabic{table}}

\section*{Supporting Information}
\label{support}

\subsection*{S1. Moist-air speed-of-sound calculation}

ForestIR computes the speed of sound from the configured atmospheric state rather than using a fixed constant. The implementation uses a lightweight ideal-mixture approximation for humid air, following standard thermodynamic treatments of sound speed in moist air \citep{Cramer1993,GaviosoEtAl2025JPR}. For an ideal gas mixture, the speed of sound can be expressed as
\begin{equation}
c \;=\; \sqrt{\gamma \, R_{\mathrm{spec}} \, T},
\label{eq:sos_ideal}
\end{equation}
where $T$ is the absolute temperature, $R_{\mathrm{spec}}$ is the mixture-specific gas constant, and $\gamma=c_p/c_v$ is the ratio of specific heats, where $c_p$ is the constant-pressure specific heat, and $c_v$ is the constant-volume specific heat \citep{Moser2009}. Following established thermodynamic treatments of sound speed in moist air \citep{Cramer1993,GaviosoEtAl2025JPR}, we allow both $\gamma$ and $R_{\mathrm{spec}}$ to vary with humidity and pressure through the composition of humid air.

Given temperature $T$ (K), pressure $p$ (kPa), and relative humidity $\mathrm{RH}$ (\%), we first compute the saturation vapour pressure ${p_{\mathrm{sat}}(T)}$ using the ISO~9613-1 formulation (expressed here as a ratio to the reference pressure $p_{\mathrm{ref}}=101.325$~kPa),
\begin{equation}
\frac{p_{\mathrm{sat}}(T)}{p_{\mathrm{ref}}}
=
10^{\, -6.8346\left(\frac{T_{01}}{T}\right)^{1.261} + 4.6151},
\qquad
T_{01}=273.15+0.01~\mathrm{K},
\label{eq:psat_iso}
\end{equation}
where $T_{01}$ is a fixed reference temperature appearing in the empirical saturation-vapour-pressure approximation used in ISO~9613-1 implementations, and obtain the water-vapour partial pressure $p_w = (\mathrm{RH}/100)\,p_{\mathrm{sat}}(T)$. Under the ideal-mixture approximation, the water-vapour mole fraction is then
\begin{equation}
x_w \;=\; \frac{p_w}{p}.
\label{eq:xw}
\end{equation}
Using the mole-fraction-weighted molar mass,
\begin{equation}
M_{\mathrm{mix}} \;=\; (1-x_w)M_d + x_w M_w,
\label{eq:Mm}
\end{equation}
with $M_d$ and $M_w$ denoting the molar masses of dry air and water vapour, the mixture-specific gas constant is $R_{\mathrm{spec}}=R/M_{\mathrm{mix}}$, where $R$ is the universal gas constant. 
Let $Y_w$ denote the water-vapour mass fraction implied by $x_w$,
\begin{equation}
Y_w \;=\; \frac{x_w M_w}{M_{\mathrm{mix}}}, \qquad Y_d=1-Y_w.
\label{eq:mass_fraction}
\end{equation}
We then approximate the mixture heat capacity as a mass-fraction-weighted average,
\begin{equation}
c_p \;\approx\; Y_d\,c_{p,d} + Y_w\,c_{p,w}, \qquad c_v = c_p - R_{\mathrm{spec}}, \qquad \gamma=\frac{c_p}{c_v},
\label{eq:cp_mix}
\end{equation}
where $c_{p,d}$ and $c_{p,w}$ are the mass-specific constant-pressure heat capacities of dry air and water vapour, respectively.
This yields a humidity- and pressure-dependent $c$. While \citet{Cramer1993} provides a high-precision calculation and interpolating expressions for $\gamma$ and $c$ over typical atmospheric ranges, the above mixture-based construction captures the same qualitative dependencies and is consistent with the simulator's use of temperature, humidity, and pressure in ISO~9613-1 air-absorption calculations.

\subsection*{S2. Humidity dependence of the speed of sound}

Table S1 provides a numerical check of the moist-air speed-of-sound calculation at fixed temperature and pressure. At $T=20^\circ\mathrm{C}$ and $p=101.325$~kPa, the legacy temperature-only approximation remains constant, whereas the updated calculation varies with relative humidity: it is slightly lower at RH=0\% and increases monotonically as humidity rises.

\begin{table}[h]
\centering
\caption{Example validation of humidity dependence in the speed of sound at $T=20^\circ\mathrm{C}$ and $p=101.325$~kPa.}
\label{tab:sos_validation}
\begin{tabular}{lcc}
\toprule
Relative humidity & Legacy $c(T)$ (m/s) & Updated $c(T,\mathrm{RH},p)$ (m/s) \\
\midrule
0\%   & 343.4200 & 343.2077 \\
20\%  & 343.4200 & 343.4595 \\
40\%  & 343.4200 & 343.7121 \\
60\%  & 343.4200 & 343.9654 \\
80\%  & 343.4200 & 344.2196 \\
100\% & 343.4200 & 344.4746 \\
\bottomrule
\end{tabular}
\end{table}

\subsection*{S3. Microphone Array Simulation}
To simulate realistic forest background conditions, ForestIR optionally adds additive noise to each microphone channel,
\begin{equation}
y_m[n] \;=\; \tilde{y}_m[n] \;+\; v_m[n].
\label{eq:add_noise}
\end{equation}
We implement a set of noise models $v_m[n]$ designed for controlled robustness tests as well as field-realistic masking:
(i) additive white Gaussian noise (AWGN),
(ii) pink noise (approximately $1/f$ power-law),
(iii) recorded environmental background noise segments,
(iv) Brown (random-walk) noise,
and (v) band-limited noise produced by FIR filtering. The colored-noise variants follow standard power-law noise constructions \citep{Kasdin1995}.

For synthetic noise (AWGN, pink, Brown, band-limited), ForestIR generates independent realizations per channel and normalizes each channel to unit RMS before scaling \citep{Kasdin1995}. For recorded environmental noise, we provide an example configuration using the DEMAND multichannel noise database \citep{ThiemannItoVincent2013DEMAND} and extract per-channel noise by sampling \emph{non-overlapping} segments from a single long recording (with resampling if needed), then normalizing each channel to unit RMS. This non-overlap strategy avoids introducing artificial inter-channel coherence that can unrealistically benefit localization methods.

Noise magnitude is controlled by a user parameter \texttt{noise\_level} that maps to a fixed absolute noise gain (in dBFS), independent of the clean signal amplitude. Concretely, ForestIR first generates a ``unit'' noise realization and then rescales it to the target level; this design makes the resulting SNR vary across microphones according to geometry and propagation, matching the fact in reality that for one source sound, different propagation distances would lead to different SNR level at different sensors.

\subsection*{S4. ForestIR Code Configuration}

\begin{sloppypar}

\paragraph{Interface and execution modes.}
ForestIR provides a batch command-line interface and a programmatic Python API, both built on the same simulation core in \texttt{code/ForestReverb.py}.
Batch rendering iterates over all rows in a source-location CSV and, for each scene, produces (i) a multichannel mixture WAV, (ii) separate mono per-microphone WAVs (written to a \texttt{per\_mic\_\{tag\}/} subdirectory), and (iii) one row in a CSV manifest (\texttt{manifest.csv}) recording metadata for that render.
Batch mode is invoked as
\begin{verbatim}
python code/render_batch.py --bird-wav <path/to/clean/bird.wav> [options]
\end{verbatim}
Relative file paths are resolved from the repository root; bundled presets live under \texttt{dataset/}.
Alternatively, the modules in \texttt{code/} expose the same underlying functions---e.g.,
\texttt{simulateForestIR} (forest IR synthesis),
\texttt{render\_array\_from\_source} (source--IR convolution),
and \texttt{synthesize\_array\_observation} (end-to-end render with optional noise and WAV export)---for single-scene rendering or custom pipelines.
The programmatic API does not invoke the CLI and does not automatically perform batch iteration or write \texttt{manifest.csv} unless the user implements that loop.
Dependencies are listed in \texttt{requirements.txt} (\texttt{pip install -r requirements.txt}).

\paragraph{Inputs and configuration.}
A ForestIR run is specified by (i) source audio and positions, (ii) array geometry, (iii) forest structure, and (iv) environmental and noise settings.

\textit{Source audio and positions.}
The batch CLI requires a clean source WAV (\texttt{--bird-wav}), which is convolved with the simulated array impulse response at each source location.
Source positions are read from a CSV with columns \texttt{x,y,z} (metres).
Built-in layouts are selected with \texttt{--src} (\texttt{custom} [default], \texttt{finland\_konnevesi}, or \texttt{finland\_konginkangas}), or overridden with a user CSV via \texttt{--src-csv}.

\textit{Array geometry.}
Array geometry is set with \texttt{--mic} or \texttt{--mic-csv}.
Built-in microphone layouts correspond to field deployments:
\texttt{finland\_konnevesi} (Konnevesi BPH array),
\texttt{finland\_konginkangas\_11}, and
\texttt{finland\_konginkangas\_11\_1.2m} (Konginkangas array, optionally filtered to $z=1.2$\,m).
Custom layouts use a CSV with header \texttt{device,x,y,z}; row order defines channel order, and \texttt{device} labels per-channel output files (e.g., \texttt{F3}--\texttt{F13} or \texttt{BPH1}--\texttt{BPH3}).

\textit{Forest structure.}
Forest structure is configured with \texttt{--tree-csv}:
measured tree locations (\texttt{default} or \texttt{finland}, mapping to \texttt{dataset/treeloc\_finland.csv}),
a user-provided tree CSV,
a treeless scene (\texttt{none}),
or a synthetic forest (\texttt{random} with \texttt{--nTrees} $> 0$).
For \texttt{random} trees, sampling is controlled by \texttt{--tree-sampling-algo} (\texttt{uniform} or \texttt{repulsive}) and \texttt{--min-tree-spacing}.
Trunk geometry uses \texttt{--tree-height} and a trunk-diameter range (\texttt{--trunk-diameter-min/max}).
Optional branch/leaf scattering (\texttt{--apply-branch-leaf}) adds diffuse scatterers whose count and vertical bounds are set by \texttt{--n-branch-leaf-scatterers} and \texttt{--branch-leaf-z-min/max}, with horizontal extent from \texttt{--forest-range-x/y}.
Ground reflection on the floor-reflection path is set by preset (\texttt{--ground-type}: \texttt{default}, \texttt{concrete}, \texttt{water}, \texttt{ice}, \texttt{grass}, \texttt{snow}) or by a manual coefficient (\texttt{--floor-reflection-coef}).

\textit{Environment, sampling, and noise.}
Environmental parameters---temperature (\texttt{--temp-c}), relative humidity (\texttt{--rh-percent}), and pressure (\texttt{--pressure-hpa})---and the output sampling rate (\texttt{--fs}; default 384\,kHz) are exposed as CLI arguments.
Reproducibility of stochastic components uses two seeds:
\texttt{--scene-seed} governs geometric randomness (random tree placement, trunk parameters, and branch/leaf scatterer placement), and
\texttt{--noise-seed} governs additive noise (in batch mode, source index~$i$ uses \texttt{noise\_seed + \textit{i}}).
Background noise is added after the clean render via selectable models (\texttt{white}, \texttt{pink}, \texttt{real\_forest} [recorded environmental noise], \texttt{brown}, or \texttt{bandlimited}) with level control (\texttt{--noise-level}), defined as the target ratio $\mathrm{RMS}(\text{noise})/\mathrm{RMS}(\text{signal})$ after rendering ($0$ disables added noise).
For \texttt{real\_forest}, recorded-noise presets (\texttt{--noise-source nriver} or \texttt{konnevesi}) or an explicit noise-audio path or directory (\texttt{--noise-path}) may be supplied.

\paragraph{Outputs and reproducibility.}
ForestIR batch mode produces convolved array \emph{recordings} rather than exported impulse-response WAV files: for each source location, (i) one multichannel WAV, (ii) mono per-microphone WAVs, and (iii) a manifest row linking outputs to the full experimental configuration.
The manifest stores output paths (\texttt{mc\_wav}, \texttt{per\_mic\_dir}), a stable source identifier (\texttt{src\_uid}), source coordinates, microphone/source preset identifiers and CSV paths, the input bird WAV, forest configuration (\texttt{tree\_mode}, tree CSV or random-sampling settings, branch/leaf scattering toggle), environmental parameters, ground-reflection settings, the chosen noise model and its parameters, and both \texttt{scene\_seed} and the per-source \texttt{noise\_seed}.
This design supports systematic benchmarking of localization pipelines and scalable synthetic training-data generation while preserving full provenance of each rendered example.
\end{sloppypar}

\end{document}